\documentclass[12pt]{article}

\usepackage{amssymb,amsmath,eucal}%,epsf,graphicx,epic}
\newcommand{\e}{\epsilon}
\newcommand{\f}{\varphi}

\newcommand{\s}{\sigma}

\newcommand{\F}{\Phi}

\newcommand{\La}{\Lambda}

\font\largerm=cmr12 at 24pt

\font\cmssl=cmss10 at 12 pt

\font\cmsslll=cmss10 at 14 pt

\input cyracc.def
\font\tencyr=wncyr10
\def\cyr{\tencyr\cyracc}

\newcommand{\bR}{\mathbb{R}}
\newcommand{\bZ}{\mathbb{Z}}

\newcommand{\bO}{\mathbb{O}}

\renewcommand{\gg}{\mathfrak{g}}

\newcommand{\gu}{\mathfrak{u}}

\newcommand{\so}{\mathfrak{so}}
\newcommand{\su}{\mathfrak{su}}
\newcommand{\spin}{\mathfrak{spin}}

\newcommand{\cI}{\mathcal{I}}

\newcommand{\cP}{\mathcal{P}}

\newcommand\fr[2]{\tfrac{#1}{#2}}

\newcommand\LR[2]{\Lambda^{#1}\mathbb{R}^{#2}}

\newcommand{\ra}{\rightarrow}

\DeclareMathOperator\id{id}

\newcommand{\wh}{\widehat}
\newcommand{\ol}{\overline}
\newtheorem{Th}{Theorem}
\newtheorem{Prop}{Proposition}
\newtheorem{Cor}{Corollary}
\newtheorem{Lem}{Lemma}
\newtheorem{Def}{Definition}
\newtheorem{Ex}{Example}[subsection]
\newtheorem{Pres}{Presentation}
\newtheorem{Conj}{Conjecture}

\newcommand{\bpr}{\begin{Pres}\ \ }
\newcommand{\epr}{\end{Pres}\ \ }
\newcommand{\bex}{\begin{Ex}\ \ }
\newcommand{\eex}{\end{Ex}\ \ }
\newcommand{\bt}{\begin{Th}\ \ }
\newcommand{\et}{\end{Th}}
\newcommand{\bp}{\begin{Prop}\ \ }
\newcommand{\ep}{\end{Prop}}
\newcommand{\bc}{\begin{Cor}\ \ }
\newcommand{\ec}{\end{Cor}}
\newcommand{\bl}{\begin{Lem}\ \ }
\newcommand{\el}{\end{Lem}}
\newcommand{\bd}{\begin{Def}\ \ }
\newcommand{\ed}{\end{Def}}
\newcommand{\bconj}{\begin{Conj}\ \ }
\newcommand{\econj}{\end{Conj}}

\catcode`@=11
\@addtoreset{equation}{section}
\catcode`@=12

\newcommand{\n}{\nabla}
\newcommand{\op}{\oplus}
\newcommand{\ot}{\otimes}

\newcommand{\be}{\begin{equation}}
\newcommand{\ee}{\end{equation}}
\newcommand\la[1]{\label{#1}}
\newcommand\re[1]{(\ref{#1})}
\def\<#1,#2>{\langle\,#1,\,#2\,\rangle}
\newcommand{\arr}{\begin{array}{rlll}}
\newcommand{\ea}{\end{array}}
\newcommand{\bea}{\begin{eqnarray}}
\newcommand{\eea}{\end{eqnarray}}
\newcommand{\bean}{\begin{eqnarray*}}
\newcommand{\eean}{\end{eqnarray*}}

\numberwithin{equation}{section}
\newcounter{ssig}
\newcommand{\sig}{\refstepcounter{ssig}\thessig}
\newcounter{ttig}
\newcommand{\tig}{\refstepcounter{ttig}\thettig}
\begin{document}
\begin{titlepage}
%\rightline{Matr11,\today}
\vskip 1 cm
\begin{center}
{\largerm   Matryoshka of Special Democratic Forms}
\vskip 1 true cm
{\cmsslll Chandrashekar Devchand$^a$, Jean Nuyts$^b$ and Gregor Weingart$^c$ }
\vskip 0.8 true cm
{
\small  devchand@math.uni-potsdam.de\ ,\ jean.nuyts@umh.ac.be\ , gw@matcuer.unam.mx 
\vskip 0.2 true cm
{\it $^a$ Institut f\"ur Mathematik der Universit\"at Potsdam,}
\\[1pt]
{\it Am Neuen Palais 10, D-14469 Potsdam, Germany}
\\[2pt]
{\it $^b$  Physique Th\'eorique et Math\'ematique,
Universit\'e de Mons-Hainaut}
\\[1pt]
{\it 20 Place du Parc, B-7000 Mons, Belgium}
\\[1pt]
{\it $^c$  Instituto de Matem\'aticas, Universidad Nacional Aut\'onoma de M\'exico}
\\[1pt]
{\it 62210 Cuernavaca, Morelos, Mexico}
}
\end{center}
\vskip 1 true cm
\begin{abstract}
Special $p$-forms are forms which have components
$\varphi_{\mu_1\dots\mu_p}$ equal to $+1,-1$ or $0$ in some orthonormal basis.
A  $p$-form $\varphi\in \Lambda^p\bR^d$ is called democratic if the set of
nonzero components $\{\varphi_{\mu_1\dots\mu_p}\}$ is symmetric under 
the transitive action of a subgroup of O$(d,\bZ)$ on the indices $\{1,\dots,d\}$. 
Knowledge of these symmetry groups
allows us to define mappings of special democratic $p$-forms  in 
$d$ dimensions to  special democratic $P$-forms in $D$ dimensions
for successively higher $P \geq p$ and $D \geq d$. In particular, we display a
remarkable nested stucture of special forms including a U(3)-invariant 2-form 
in six dimensions, a G$_2$-invariant 3-form in seven dimensions, 
a Spin(7)-invariant 4-form in eight dimensions and a  special democratic 
6-form $\Omega$ in ten dimensions. The latter has the remarkable property that 
its contraction with one of five distinct bivectors, yields,  
in the orthogonal eight dimensions, the Spin(7)-invariant 4-form. 
We discuss  various properties of this ten dimensional form.  
\end{abstract}
\end{titlepage}
\vfill\newpage

\section{Introduction}

Special holonomy plays an important role in field theories. 
For instance, supersymmetry often requires that target manifolds have
special holonomy. This property is also important for Yang-Mills theories.  
In dimensions greater  than four, 
special holonomy offers the possibility of constructing solutions of the 
Yang-Mills equations satisfying the generalised self-duality equations 
first introduced for flat Euclidean spaces in \cite{cdfn} (see also \cite{fn,bdn,dn});
\begin{equation}
\fr12   T_{mnpq}F_{pq} = \lambda F_{mn}, \quad m,n,\dots= 1,\dots,d\ .
\label{tdual}
\end{equation}
Here,  $F_{mn}$ are components of the curvature of a Yang-Mills connection $\n$ 
taking values in the Lie algebra of the gauge group and $T_{mnpq}$ are
components  of a 4-form $T$.
This 4-form acts as an endomorphism on the space of 2-forms. The curvature
$F$ restricted to eigenspaces of $T$ corresponding to nonzero eigenvalues $\lambda$
satisfy the Yang-Mills equations $\n_m F_{mn}=0$
in virtue of the Bianchi identities  $\n_{[m} F_{np]}\equiv 0$.
Interesting examples are the 4-forms invariant under 
(Sp($n$)$\otimes$Sp($1$))/$\bZ_2$, Spin(7) and G$_2$ corresponding to Yang-Mills 
equations on quaternionic K\"ahler and exceptional holonomy manifolds 
(see e.g. \cite{w,funi,cs,n,t,acd}).
Further examples of special holonomy structures are the U($n$) invariant K\"ahler 2-forms in 
$2n$ (real) dimensions and the G$_2$ invariant Cayley 3-form in seven dimensions.  
It turns out that the latter forms are not only related to each other, but also to 
interesting higher rank forms in higher dimensions.

Recall that a constant $p$-form $\f$ in a $d$-dimensional
Euclidean space is a calibration if for any $p$-dimensional subspace
spanned by a set of orthonormalised
vectors $e_1,\dots, e_p$,
\begin{equation}
(\f(e_1,\dots, e_p))^2 \le 1\ ,
\la{cal0}
\end{equation}
where equality holds for at least one subspace.
Constant p-forms can always be rescaled to be calibrations.
Many of the interesting calibrations which characterise special holonomy 
manifolds can be presented as special forms, all of whose nonzero
components saturate the bound \re{cal0} (see Definition \ref{spdef}).

In this article, we wish to highlight relationships between 
special $p$-forms in $d$ dimensions and certain special
$P$-forms ($P\ge p$) in $D$
dimensions  ($D\ge d$), governed by discrete symmetries. 
Symmetric ways
of embedding the $d$-dimensional space in the $D$-dimensional space leads us 
to a notion of democratic forms. We study examples, focusing our attention on 
specially interesting structures in dimensions seven, eight and ten.  
A remarkable nested structure, reminiscent of a 
matryoshka\footnote[1]{{\cyr matr\"eshka}, a nested Russian doll.}, 
emerges in successively higher dimensions.  In particular, this structure 
provides new examples of self-dualities.

\section{Special forms, symmetries and democracy}

We  concentrate on what we call {\it special forms}.
Let $(e_1,\dots, e_d)$ denote an orthonormal basis of $\bR^d$.
\bd\label{spdef}
 A {\cmssl special p-form} $\f$ is a $p$-form $\f\,\in\,\La^p\bR^d$ on
 $d$--dimensional Euclidian space $\bR^d$ in the orbit under the special
 orthogonal group $\mathrm{SO}(d,\bR)$ of
 \begin{equation}
  \f\;\;=\;\;\sum_{1\leq \mu_1<\ldots<\mu_p\leq d}
  \f_{\mu_1\ldots\mu_p}\,e_{\mu_1}\wedge e_{\mu_2}\wedge\ldots\wedge e_{\mu_p}
 \la{special1}
 \end{equation}
 with $\f_{\mu_1\ldots\mu_p}\,\in\,\{-1,0,1\}$.
\ed 
Hence, a $p$-form $\f$ is special if there exist $d$ orthonormal basis vectors 
$ e_\mu, \,\mu=1,\ldots,d$ such that for any subset of $p$  vectors
$e_{\mu_1},\ldots, e_{\mu_p}$ we have
\begin{equation}
 \f_{\mu_1\ldots\mu_p} \;\;=\;\;
 \f(e_{\mu_1},\ldots, e_{\mu_p}) \;\;\in\;\; \{-1,0,1\}.
 \la{special2}
\end{equation}

Trivial examples are the volume forms in $d$ dimensions, which provide the 
Hodge-duality operators mapping $p$-forms to $(d{-}p)$-forms. 
Further well-known examples are the G$_2$-invariant Cayley 3-form in seven dimensions
defined by the structure constants of the octonions, and the Spin(7)-invariant 
4-forms in eight dimensions (see sections \ref{g2sect} and \ref{spin7sect}).

The orbits of special $p$-forms under $\mathrm{SO}(d,\bR)$ or 
$\mathrm{O}(d,\bR)$ play a major role in the following.
Clearly there are only a finite number of orbits of special $p$-forms 
parametrised by the components $\f_{\mu_1\ldots\mu_p}\,
\in \,\{-1,0,1\}$ under these two groups. \
Note  however that distinct sets of components may give rise to
special $p$-forms in the same orbit, because the
subgroups $\mathrm{SO}(d,\bZ)\,\subset\,\mathrm{SO}(d,\bR)$ or 
$\mathrm{O}(d,\bZ)\,\subset\,\mathrm{O}(d,\bR)$ 
map the special form $\f$ in equation \re{special1}
into a special form parametrised by different components. These groups
are isomorphic to the semidirect product of the permutation group $S_d$
acting naturally on $d{-}1$ or $d$ copies of $\bZ_2$, namely 
$\mathrm{SO}(d,\bZ)\,\cong\,S_d\ltimes\bZ_2^{d-1}$ or
$\mathrm{O}(d,\bZ)\,\cong\,S_d\ltimes\bZ_2^d$. 
Thus, special $p$-forms which appear to be different may nevertheless be in the 
same orbit under $\mathrm{SO}(d,\bR)$ or $\mathrm{O}(d,\bR)$. 
The action of an element $(\s,\eta_1,\dots,\eta_d)\in S_d\ltimes\bZ_2^d$,
on the components of $\f$ is given by
\begin{equation}
%S_d\ltimes\bZ_2^d \ni (\s,\eta_1,\dots,\eta_d) \colon
\f_{i_1\ \dots\ i_p} \mapsto \eta_{i_1}
\dots\eta_{i_p}\,\f_{\s(i_1)\ \dots\ \s(i_p)}\ ,
\label{tsfOd}
\end{equation}
where 
$\eta_i^2=1\,,\,i=1,\dots,d$. 
If $\pi(\sigma)$ is the parity of $\sigma$, the elements 
of the subgroup 
$\mathrm{SO}(d,\bZ)\subset \mathrm{O}(d,\bZ) $ 
are those with $\eta_1 \eta_2\dots \eta_d \pi(\sigma)=1$.
The orbit of a special $p$-form may always be labeled by 
a choice of a representative \re{special2}.

If a $p$-form $\f$ is special, then $-\f$ is obviously also special
as is its Hodge dual $(d{-}p)$-form ${\star}\f$. 
For $p$ odd, $-\f$ is always in the $\mathrm{O}(d,\bR)$ orbit of
$\f$; for instance using the parity transformation $e_j\mapsto {-}e_j, \ j=1,\dots,d$.
We note that forms which can be brought to the special form \re{special1} 
by a rescaling are also interesting.

An alternative description of special forms was given in \cite{dnw}. 
Oriented sets were defined as equivalence classes of finite, totally ordered sets
up to even permutations, i.e.\ every set has two different orientations
differing by a ``sign''. Thus the oriented subsets of $\{1,\ldots,d\}$ are
in bijective correspondence to oriented coordinate subspaces $\bR^p\,\subset\,
\bR^d$ via 
$s= \{\mu_1,\ldots,\mu_p\}\,
\longmapsto\,e_{\mu_1}\wedge\ldots\wedge e_{\mu_p}$. 
A special $p$-form can be thought of as a function from the
set of oriented subsets $\{\mu_1,\ldots,\mu_p\}\,\subset\,\{1,\ldots,d\}$ to
$\f_{\mu_1\ldots\mu_p}\,\in\,\{-1,0,1\}$ with the property that the function's
values on the two different orientations of the same subset differ by a sign.
Consequently a special $p$-form is specified completely by either of the
two sets $\cI^\pm$ of oriented subsets $\{\mu_1,\dots,\mu_p\}\,\subset\,
\{1,\dots,d\}$ with $\f_{\mu_1,\ldots,\mu_p}\,=\,\pm1$. We will call the
set $\cI\,:=\,\cI^+\cup\cI^-$ the support, $|\f|\,:=\,\fr12|\cI|$ the
weight of $\f$. We denote by 
$\mu^{(a)}:= \{\mu^{(a)}_1,\dots,\mu^{(a)}_p\}, a=1,\dots,|\f|$, 
the elements in $\cI^+$, i.e.
\begin{equation}
\f = \sum_{a=1}^{|\f|} \  
% \f_{\mu_1^{(a)}\dots \mu_p^{(a)}} \ 
                       e_{\mu_1^{(a)}}\wedge \dots\wedge e_{\mu_p^{(a)}}\ .
\la{mua}
\end{equation}
Restricted to the $p$-dimensional subspace spanned by any
$\{\mu_1,\dots,\mu_p\}$ belonging to the set $\cI^+$, the
form $\f$ is equal to the $p$-dimensional volume form
and hence its components  are
 \begin{equation}
\f_{\mu_1\dots\mu_p}
 = \e_{\mu_1\dots\mu_p}
\la{special4}
\end{equation}
where $\e$ is the completely antisymmetric tensor with $p$ indices.
This means in particular that a special form $\f$ in $d$ dimensions
has non-zero components given by
\begin{equation}
\f_{\s(1)\s(2)\ \dots\ \s(p)}= \e_{12\  \dots\ p}
\quad {\rm for\ all }\quad \s\in H
\la{cal1}
\end{equation}
where $H$ is an appropriate set of permutations $\sigma$
of the indices $\{1,\dots,d\}$
which map $\{1,\dots,p\}$ to $\{\mu_1,\dots,\mu_p\}\in \cI^+$.

We can define a metric on the {\it vertex space} $\cP^p(d)$,
the space of (unoriented) $p$--element subsets of $\{1,\dots,d\}$, by setting 
$\mathrm{dist} (s,\tilde s)\,=\,p-|s\cap\tilde s|$ and visualise
the restriction of this metric to the set $\cI^+$ by drawing a graph with
labeled edges, the vertices correspond to the elements of $\cI^+$, the
edges run between vertices of distance strictly less than $p$ and are
labeled by this distance \cite{dnw}. 
The graph of a special $p$-form
$\f$ does not  completely specify the components
$\f_{\mu_1\dots\mu_p}\,\in\,\{-1,0,1\}$ up to
the action of $\mathrm{O}(d,\bZ)$; we still need to specify some relative
sign. Nevertheless the graph gives a very condensed way of encoding the
characteristics of a special $p$-form. In particular it is a useful tool
in calculating the bisymmetry group introduced below.

\begin{itemize}

\item
We call a  $p$-form $\f$ {\it permutation symmetric} under the action of 
some element $\s\,\in\,S_d$ if 
\begin{equation}
\f_{\s(i_1)\,\dots\,\,\s(i_p)}\;\;=\;\; \kappa \f_{i_1\,\dots\,\, i_p}\ ,\quad 
 \{i_1,\dots, i_p\}  \in \cI^+ ,
\la{permsymm}
\end{equation}
with $\kappa{=}1$. 
The set of all such transformations is the 
{\em permutation symmetry group} of $\f$, $G_r\subset S_d$, where
$r\,=\,|G_r|$, the order.
If $\kappa{=}-1$, we call the permutation $\s$ a {\it permutation antisymmetry}. 
If there exists one such antisymmetry $\tau$, then there are $r$
antisymmetries $\s\tau$ with $\s\in G_r$ and the set
$B= \{\s\,,\,\s\tau \mid \s\in G_r \}\subset S_d$
forms a group of order $2r$, which we call the {\it permutation bisymmetry group}.
The permutation symmetry group $G_r$ is an invariant subgroup and $B /G_r=\bZ_2$.

\item
We call a $p$-form  {\it orthogonal symmetric}  under the action of 
an element \linebreak $(\s,\eta_1,\dots,\eta_d)\,\in\,S_d\ltimes\bZ_2^d$ if
\begin{equation}
\f_{\s(i_1)\,\dots\,\,\s(i_p)}\;\;=\;\;
\kappa\eta_{i_1}\,\dots\,\eta_{i_p}     \f_{i_1\,\dots\,\, i_p}\ ,\quad 
 \{i_1,\dots, i_p\}  \in \cI^+ ,
\la{orthsymm}
\end{equation}
with $\kappa{=}1$. 
The set of all such transformations forms a group called the {\em orthogonal
symmetry group of $\f$},  denoted by $\wh G_s\subset S_d\ltimes\bZ_2^d$ 
where   $s\,=\,|\wh G_s|$.
 
If $\kappa{=}-1$ 
The transformation $(\s,\eta_1,\dots,\eta_d)$ is called an {\it orthogonal antisymmetry}.
The union of orthogonal symmetries and orthogonal antisymmetries will be called the 
{\it orthogonal bisymmetries}.

\item
We denote by $L^{(a)}\subset G_r$ (or $L^{(a)}\subset \widehat G_r$) the 
permutation (or respectively orthogonal) stabiliser of the oriented set
$\mu^{(a)} \in \cI^+$ 
%$\mu^{(a)}=\{\mu^{(a)}_1,\,\dots\,,\,\mu^{(a)}_p \}$
as the set of $\s \in G_r$ (or $\widehat G_r$) such that
$ \{ \s(\mu^{(a)}_1),\dots,\s(\mu^{(a)}_p) \} \equiv
 \{ \mu^{(a)}_1,\dots,\mu^{(a)}_p \}  $
up to an even permutation.
We define the
{\it permutation} (resp. {\it orthogonal}) {\it stability group} as
\begin{equation}
L =  \bigcap_{a=1}^{|\f|} L^{(a)}\ .
\la{pstabgr}
\end{equation}
\end{itemize}

\noindent
For example, the $d$-dimensional completely antisymmetric tensor
 $\e_{i_1\,\dots\,i_d}$
has the alternating group $A_d$ as its permutation symmetry group and
$B = S_d$, the symmetric group, as its permutation bisymmetry group 
which is also its orthogonal symmetry group.
The stability group $L$ of the set of indices
$\{1,2,\dots,d\}$ of its sole nonzero component is $A_d$ itself.
As a consequence, for a form $\f$, a symmetry (resp. antisymmetry)  
which is an even permutation in $S_d$ corresponds to a symmetry
(resp. antisymmetry) of ${\star}\f$.
On the other hand, a symmetry (resp. antisymmetry) of  $\f$ which
is an odd permutation  corresponds to an antisymmetry
(resp. symmetry) of ${\star}\f$.

\bd
A p-form with components $\f_{i_1 i_2 \dots i_p}$ is called 
{\cmssl democratic} if there is a transitive action of one of the 
permutation bisymmetry groups on $\{1,\dots,d\}$. In other words,
for any $\,i_1,i_2\in \{1,\dots,d\}$ there exists at least 
one element $\s$ of one of the 
permutation bisymmetry groups such that $\s(i_1)=i_2 $.
\ed
Hodge-duality is clearly a bijection amongst democratic forms.  

The interplay between two related $p$- and $P$-forms in dimensions $d$ and $D$
respectively and the connections between their related symmetry and bisymmetry
groups turns out to be an interesting subject of study.
We shall use these in order to discuss the `presentations' of some particularly
interesting special forms and also to construct $P$-forms on $D$-manifolds
from $p$-forms ($p\leq P$) on $d$-manifolds ($d<D$).

\section{Higher dimensional forms from lower dimensional forms }

We consider interesting ways of constructing $P$-forms $\F$ in a
$D$--dimensional space $\bR^D$ from $p$-forms $\f$ on a $d$--dimensional
subspace $\bR^d$ of $\bR^D$, with $P=p+b\,,\,D=d+a$ and $0\leq b\leq a$.
Let us embed $\bR^d$ in $\bR^D$  in such a way that the
first $d$ basis vectors of $\bR^D$ are identical to the basis vectors
$e_i\,,\, i=1,\dots,d$ of $\bR^d$. We label the basis vectors of $\bR^D$
by  indices  
$i=1\, ,\, \dots\, ,\, d\, ,\, d{+}1\, ,\, \dots\, ,\, D{-}b,\, D{-}b{+}1\, ,\, \dots\, ,\, D$.

Given the components of a $p$-forms $\f\in \La^p\bR^d$ 
and an appropriate subgroup $H\subset S_D$, we may define
a $P$-form $\F\in \La^P\bR^D$ with nonzero components given by
\begin{equation}
\left\{ \F_{\Sigma(i_1)\,\dots\,\Sigma(i_p)\Sigma(D-b+1)\,\dots\,\Sigma(D)}
= \f_{i_1\,\dots\,i_p}
\mid
\ i_1,\dots,i_p \in \{1,\dots,d\}\ ,\  \Sigma\in H 
\right\},
\la{qd}
\end{equation}
where the $\Sigma\in H$  satisfy the following 

 \begin{description}
  \item{Compatibility condition:}
  
 Consider any two sets of indices
$ \{ i_1 ,\dots, i_p \}$  and $\{ j_1 ,\dots, j_p \}$
belonging to $\{1,\dots,d\}$
such that
      $ \f_{ i_1 \dots i_p } = \f_{ j_1 \dots j_p } $.
If two permutation $\Sigma,\Sigma' \in H$  have the property that
\bea
&&\{ \Sigma(i_1)  ,\dots, \Sigma(i_p) ,\Sigma(D{-}b{+}1),\dots,\Sigma(D)\}
\nonumber\\
&=& \{ \Sigma'(j_1) ,\dots, \Sigma'(j_p),\Sigma'(D{-}b{+}1),\dots,\Sigma'(D) \}
\la{comp1}
\eea
then $\Sigma,\Sigma'$ must be  compatible in the sense that
\begin{equation}
  \F_{ \Sigma(i_1) \dots \Sigma(i_p) \Sigma(D-b+1) \dots \Sigma(D) }
 = \F_{ \Sigma'(j_1) \dots \Sigma'(j_p) \Sigma'(D-b+1) \dots \Sigma'(D) }\ .
\la{comp2}
\end{equation} 
\end{description}
It is clear that the restriction of $\F$ to the $d$-dimensional subspace yields 
$\f$, i.e. $\F\vert_{\bR^d}=\f$, since the identity obviously belongs to $H$.

As we shall see in the next section, interesting cases arise when the forms
are special and the subgroup $H$ is chosen so as to ensure 
democracy amongst the indices of $\F$. Particularly interesting examples
correspond to the following three restrictions of \re{qd}.

\begin{description}
\item{\bf{A: }} For $D=d\ ,\ P=p$, 
consider the non-zero components of a $p$-form  $\F\in \La^p\bR^D$, 
with discrete symmetry group $G_r$ (see \re{permsymm}).
These are generated as the orbit
under some subgroup $H \subset G_r$ of a set of  $|\f| \le |\F|$ 
given non-zero components, 
$\f_{\mu^{(a)}_1\,\dots\,\mu^{(a)}_p }\ ,\ a=1,\dots,|\f|$, which define $\F$ thus: 
\begin{equation}
\F_{_{\sigma(\mu^{(a)}_1)\,\dots\,\,\sigma(\mu^{(a)}_p) }} =
\f_{_{\mu^{(a)}_1\,\dots\,\,\mu^{(a)}_p }}
 \ ,\ 
 a=1,\dots,|\f|   \,,\,
 \sigma\in H \subset G_r\ .
\la{qa}
\end{equation}
In other words, all the components of $\F$ arise from this formula.
A {\it presentation} $P[m;n](\F)$ of the $p$-form $\F$ is defined as the set of
components $\f_{\mu^{(a)}_1\,\dots\,\,\mu^{(a)}_p} =1 $, $a=1,\dots,m {:=} |\f|$
together with a set of $n$ permutations generating a presentation of $H$.

\item{\bf{B: }} For $D> d\ ,\ P=p$, a $p$-form  $\f$ on $\bR^d$
defines a $p$-form  $\F$ on $\bR^D$ with components
\begin{equation}
\left\{
\F_{\Sigma(i_1)\,\dots\, \Sigma(i_p)} = \f_{i_1\,\dots\, i_p}
\mid
 i_1,\dots,i_p \in \{1,\dots,d\}  ,\
 \Sigma\in H 
\right\},
\la{qb}
\end{equation}
where  $H$ is some subgroup of $S_D$.

\item{\bf{C: }} For $D=d+q\ ,\  P=p+q$, a $p$-form  $\f$ on $\bR^d$
defines a $(p{+}q)$-form  $\F$ on $\bR^{d+q}$ with components
\begin{equation}
\left\{
\F_{\Sigma(i_1)\,\dots\,\Sigma(i_p)\Sigma(d+1)\,\dots\,\Sigma(D)} = \f_{i_1\,\dots\,i_p}
\mid
 i_1,\dots,i_p\in \{1,\dots,d\} ,\ 
\Sigma\in H  
\right\},
\la{qc}
\end{equation}
where  $H$ is some subgroup of $S_D$.
\end{description}

\section{Examples}
Numerous examples of the constructions in \re{qa},\re{qb} and \re{qc}
can be generated, many rather trivial, e.g.~the extension
of a 1-form in one dimension to a 2-form in two dimensions.
However, relationships between $p$-forms invariant under
the following subalgebras of $\so(D)$ provide interesting examples:
 \begin{description}
\item{a)} $\so(d)\subset \so(D)\ ,\ d<D$
\item{b)} $\su(n)\op\gu(1)\subset \so(2n)$
\item{c)} $\gg_2\subset \so(7)$
\item{d)} $\spin(7)\subset \so(8)$
\end{description}
In particular these examples fall into a remarkable nested structure of forms
in successively higher dimensions up to eight. Furthermore, this matryoshka 
extends to interesting examples of special forms in ten dimensions.

\subsection{$\so(d)$-invariant forms}
The $\so(d)$-invariant $d$-forms clearly provide completely trivial
examples. The unique nonzero component
$\e_{1\dots d}\,=\,1$, together with $\id{\in}S_d$ provides a presentation.
Moreover the $\so(d)$--invariant $d$-form $\e$ can be extended to the
$\so(D)$-invariant $D$-form {\Large$\e$} for $D>d$ as follows:
\begin{equation}
\mbox{\Large$\e$}_{\Sigma(i_1)\,\dots\,\Sigma(i_d)\Sigma(d+1)\,\dots\,\Sigma(D)}
= \e_{i_1\,\dots\,i_d} \quad,\quad \Sigma=\id.
\la{epsilon}
\end{equation}
Clearly, the $d$-dimensional completely antisymmetric tensor
 $\e_{i_1\,\dots\,i_d}$ is fully democratic, the bisymmetry group
 $S_d$ acting transitively on its indices.

\subsection{$\gu(n)$-invariant forms}

Consider the $\gu(n)$-invariant  2-form  $\omega$ on $\bR^{2n}$, 
with non-zero components
\begin{equation}
\omega_{12}=\omega_{34}=\omega_{56}=\dots =\omega_{(2n-1)(2n)}= 1\ .
\label{h6}
\end{equation} 
The vertex space $\cP^2(2n)$ consists of $n$ points. 
The  permutation symmetry group of the 2-form $\omega$  is $ G_{n!} = S_n\,$,
the group of permutations of the $n$ ordered pairs of indices,
$ \{1,2\} , \{3,4\} , \{5,6\} ,\dots, \{(2n-1),(2n)\}$, generated by
\bea
\s_{\sig\la{s26}} &:=& (\,1\ 3\,)(\,2\ 4\,)(\,5\,)(\,6\,)\dots(\,2n\,)
\nonumber\\
\s_{\sig\la{s36}} &:=& (\,1\ 3\ 5\ \dots\ 2n{-}1\,)(\,2\ 4\ 6\ \dots\ 2n\,)\ .
\la{symm6}
\eea
The number of permutation antisymmetries is also $n!$,
generated by the composition of the 
generators in \re{symm6} with, for example, the permutation
\bea
\tau_{\tig\la{t26}} := (\,1\ 2\,)(\,3\ 4\,)(\,5\ 6\,)\ \dots\ (\,2n{-}1\ \ 2n\,)\
.
\la{anti26}
\eea
There are $2^{2n} n!$ orthogonal symmetries and $2^{2n} n!$ 
orthogonal antisymmetries. These are either permutation 
symmetries or permutation antisymmetries multiplied by 
$2^{n}$ possible sign factors.

The 2-form $\omega$, as well as the  $2k$-forms $\f=\fr{1}{k!}\,\omega^k$, $k\leq n$,
are democratic and special. For example, the permutation symmetry group of the
4-form $\fr{1}{2}\,\omega^2$ is the permutation bisymmetry group of $\omega$ 
and it has no antisymmetries.

\bex
Starting from any one non-zero component, say $\omega_{12}$, 
the  other components in \re{h6} can be generated by the subgroup 
$ H_n\subset  G_{n!}$
generated by $\s_{\ref{s36}}$. Hence a presentation is given by
$P[1;1](\omega)= \{\omega_{12}\,;\,  \s_{\ref{s36}} \}$.
For $n=3$ the invariant subgroup $H_3$ is the commutator subgroup
(the closure of
the set of elements of the form $b^{-1}a^{-1}ba\ \forall a,b\in  G_{6}$).
\eex
\bex
The 2-form $\omega\in\LR{2}{2n}$  may be constructed from the 2-form in two 
dimensions, $\e\in\LR{2}{2}$, with non-zero component $\e_{12}=1$  thus:
\begin{equation}
\omega_{\Sigma(a)\Sigma(b)} = \e_{ab}\ ,\  a,b=1,2\ ,\
       \Sigma\in\{\s_{\ref{s36}}^m \mid  m=1,\dots,n \}= H_n \subset G_{n!}\ ,
\la{etoh}
\end{equation}
where $H_n$ is the subgroup generated by $\s_{\ref{s36}}$ in \re{symm6}.
\eex

\subsection{The $\gg_2$-invariant form}
\la{g2sect}

Consider the G$_2$ invariant special 3-form  $\psi$  on $\bR^7\ $,
with non-zero components  $\psi_{abc}$ given by any choice of the 
structure constants of the imaginary octonions. 
Let $\{e_a, a=1,\dots 7\}$ denote the standard
basis for Im$(\bO)\simeq \bR^7$, with   $e_a e_b = \psi_{abc}e_c - \delta_{ab}$.
A choice of the structure constants $\psi_{abc}$ is given by
\begin{equation}
     \psi_{1  2  7 }
\ =\   \psi_{1  6 3  }  \ =\   \psi_{1 5 4 }
\ =\   \psi_{2 5 3 }
\ =\  \psi_{2  4  6 }   \ =\  \psi_{3  4  7  }
\ =\  \psi_{5  6  7  }  \ = \  1\ .
\label{t7}
\end{equation}
The vertex space $\cP^3(7)$ consists of $7$ points. The 7-valent graph connecting
these vertices has all edges labeled by distance 2. The permutation symmetry group 
of $\psi$ (and naturally of its 4-form dual $\ {\star}\psi$) is a group of order
twenty-one, $ G_{21}$, generated by the permutations
\bea
\s_{\sig\la{s77}} &:=& (1\,2\,5\,4\,6\,7\,3)
\\
\s_{\sig\la{s37}} &:=& (1\,3\,5)(2\,4\,6)(7)
\eea
The permutation $\s_{\ref{s77}}$ generates the
commutator subgroup $ H_7 \subset  G_{21}$.
Since the permutation symmetry group includes an order 7
permutation, the form $\psi$ is manifestly democratic.
There are no permutation antisymmetries $\tau$. 
Using {\it Maple} we have determined that the number 
of orthogonal symmetries is 672, with an order 168 commutator subgroup
generated by either
\bea
\{(1\,5\,7)(2\,4\,6)(3),\ (1\,2\,6\,5)(4\,7)(3),\ (1\,4\,7)(2\,5\,3)(6)\}
\la{der7os}
\eea
or, alternatively, by
\bea
\{(1\,4)(2\,3)(5)(6)(7),\ (1\,2\,4\,7\,6\,3\,5),\ (1\,3\,4)(5\,6\,7)(2)\}
\la{der7as}
\eea
The number of orthogonal antisymmetries is also 672, obtained from 
the orthogonal symmetries by multiplying all the $\eta_i$ by $-1$.

\bex
The component choice \re{t7} can be generated from any non-zero
component in \re{t7} by the iterated action of $\s_{\ref{s77}}$.  
Thus a presentation is given by, for instance,
$P[1;1](\psi)=\{ \psi_{127} \,;\, \s_{\ref{s77}} \}$.
Alternatively, a less economical presentation is given by
$P[3;1](\psi)=\{ \psi_{127},\psi_{136},\psi_{246} \,;\, \s_{\ref{s37}} \}$.
\eex

\bex
The components of the 4-form dual ${}{\star}\psi\in\LR47 $ can be 
obtained in the following way:
\begin{equation}
{}{\star}\psi_{\Sigma(1)\Sigma(2)\Sigma(3)\Sigma(4)} = \e_{1234} \ ,\ 
                         \Sigma\in  H_7 \subset S_7\,
\la{C2}
\end{equation}
where $H_7$ is the group generated by the permutation
$\s_{\ref{s77}}=(1\,2\,5\,4\,6\,7\,3)$.
\eex

\bex
The 2-form $\omega$ in $\bR^6$ given by \re{h6} for $n{=}3$  
affords an extension to the 3-form $\psi$ thus:
\begin{equation}
\psi_{\Sigma(i)\Sigma(j)\Sigma(7)} = \omega_{ij} \ ,\  
                                   i,j=1,\dots,6\ ,\
                        \Sigma\in  H \subset S_7\,,
\la{htoc}
\end{equation}
where there are three types
of `minimal' choices of $H$, having only one generator:
\begin{itemize}
\item $ H= H_7$ generated by $\s_{\ref{s77}}= (1\,2\,5\,4\,6\,7\,3)$
\item  $ H= H_3$ generated by $(1)(2\,6\,5)(3\,4\,7)$
\item $ H= H_3$ generated by $(2)(1\,3\,6)(4\,7\,5)$.
\end{itemize}
\eex

\noindent
Composing the mappings in equations \re{etoh} and \re{htoc} immediately yields:
\bex
Consider the two dimensional form $\e\in\LR22\ ,\ \e_{12}=1 $.
It yields the components \re{t7} of the 3-form $\psi$ in seven dimensions:
\begin{equation}
\psi_{\Sigma(i)\Sigma(j)\Sigma(7)} = \e_{ij} \ ,\     i,j=1,2\ ,\
                        \Sigma\in  H_7 \subset S_7\,.
\la{D1}
\end{equation}
Choosing  $H_7$ to be the group generated by
$\s_{\ref{s77}}=(1\,2\,5\,4\,6\,7\,3)$
again gives the set of $\psi$'s in \re{t7}.
In fact, choosing $H_7$ to be the group generated by any seven-cycle
of the form $(1\,2\,*\,*\,*\,7\,*)$ or $(1\,2\,*\,7\,*\,*\,*)$
provides equivalent choices of components of this 3-form. 
\eex

\noindent
This example yields a simple mnemonical
construction of the structure constants of the imaginary octonions.

\subsection{The $\spin(7)$-invariant form}
\la{spin7sect}

Consider the Spin(7)-invariant self-dual 4-form
$\phi$ in $d=8$  \cite{cdfn} with non-zero components
\bea
1&=&   \phi_{1  2  3  4 } \,=\, \phi_{1  2  5  6 }\,=\, \phi_{1  2  7  8 }
\,=\, \phi_{1  3  5  7 }  \,=\,  \phi_{1 3 8 6 }  \,=\,  \phi_{1  4 8 5}
\,=\, \phi_{1  4  7  6}  \nonumber\\
&=&  \phi_{2  3  8 5}  \,=\, \phi_{2  3  7 6}  \,=\, \phi_{2  4  7 5}
\,=\, \phi_{2  4  6  8 }  \,=\,\phi_{3  4  5  6 }  \,=\, \phi_{3  4  7 8 }
\,=\, \phi_{5  6  7  8 }  .
\label{t14}
\eea
We note that each pair $(i,j)$ of indices occurs precisely thrice and the
contraction with any 2-plane spanned by $\{e_i,e_j\}$ yields the $\gu(3)$-invariant
2-form \re{h6}  with $\phi(e_i,e_j,\,\cdot\,,\cdot)\vert_{\bR^6}= \omega$.
The vertex space $\cP^4(8)$ consists of $14$ points. In the corresponding 
graph, every vertex is connected to 12 others by edges of distance 2 and to
one antipodal point at distance 4.
This form is democratic and has a permutation symmetry group $G_{168}$  
generated by 
\bea
\s_{\sig\la{s68}} &:=& (1\,2)(3\,6\,7\,4\,5\,8)
\nonumber\\
\s_{\sig\la{s78}} &:=& (8)(1\,2\,5\,4\,6\,7\,3)\ .
\la{symm168}
\eea

The permutation $\s_{\ref{s78}}$ has a $1^17^1$ cycle decomposition. Its
powers, apart from the identity, generate six independent permutations
having the same cycle decomposition. There are 8 such permutations,
corresponding to all 8 choices of the 1-cycle. In all they generate
48 permutations in the class $1^17^1$.

The permutation $\s_{\ref{s68}}$ has a $2^16^1$ cycle decomposition.
Clearly, its inverse, $\s_{\ref{s68}}^5$, also. Further,
$\s_{\ref{s68}}^2$ and  $\s_{\ref{s68}}^4$ have
$1^23^2$ cycles. There are 28 permutations of each of these four types,
corresponding to all choices of the 2-cycle in $\s_{\ref{s68}}$.
So, these generate
56 permutations in each of the classes $2^16^1$ and $1^23^2$.
The third power, $\s_{\ref{s68}}^3$, generates a $2^4$ cycle.
There are seven such
permutations. Including the identity, we therefore have the
$168=48+56+56+7+1$ elements of $ G_{168}$.

The permutation $\s_{\ref{s68}}$ clearly decomposes into the product of the
order 3 and order 2 permutations
\bea
\s_{\sig\la{s38}} &:=& (1)(2)(3\,5\,7)(4\,6\,8)
\nonumber\\
\s_{\sig\la{s28}} &:=& (1\,2)(3\,4)(5\,6)(7\,8)\ .
\la{symm56}
\eea
and a presentation for $ G_{168}$ is given by $\s_{\ref{s78}}$ and
$\s_{\ref{s38}}$.

The commutator subgroup of $ G_{168}$ is the order 56 group
generated by $\s_{\ref{s78}}$ and $\s_{\ref{s28}}$.
It contains the 48 elements in the class $1^17^1$, the seven elements in the
class $2^4$ and the identity.

The orthogonal symmetries of \re{t14} total 10752 elements,
with the commutator subgroup being the order 1344 group generated by
$(7)(1\,3\,2\,8\,4\,5\,6)$ and $(6)(1\,5\,7\,2\,8\,3\,4)$. 
The form $\phi$ has no antisymmetries. 

\bex
The order 12 subgroup $ H_{12}\subset G_{168}$ leaving the component
$\phi_{1234}$ invariant is generated by $\s_{\ref{s28}}$ and
\bea
\s_{\sig{\la{s3816}}}  :=  (1)(6)(2\,3\,4)(5\,8\,7)\ .
\la{symm12}
\eea
It has 14 left-cosets corresponding to the 14 components of $\phi_{ijkl}$,
more precisely the action of the 14 cosets on $\phi_{1234}$ generates the
14 non-zero components of $\phi$.
\eex

The 4-form $\phi$ in \re{t14} may be constructed in various ways from the 
$\so(n)$-, $\su(3)\op\gu(1)$-, and $\gg_2$-invariant forms discussed above.

\bex
Starting from the 4-form in $\bR^4$
we can generate the 4-form $\phi$ in eight dimensions with components \re{t14}
thus:
\begin{equation}
\phi_{\s(1)\s(2)\s(3)\s(4)} = \e_{1234}\ ,\ 
                         \s\in  H_{12} \subset S_8\ ,
\la{etot}
\end{equation}
where $H_{12}$ is the group generated by $\s_{\ref{s28}}$ and
$\s_{\ref{s3816}}$.
\eex

\bex
From the components \re{t7} of the $\gg_2$-invariant form $\psi\in \LR37$,
we may obtain the Spin(7)-invariant 4-form $\phi$:
\begin{equation} 
\phi_{\s(i)\s(j)\s(k)\s(8)} = \psi_{ijk} \ ,\
                  i,j,k=1,\dots,7\ ,\ 
                        \s\in  H \subset S_8\,.
\la{ctot}
\end{equation}
By choosing $H$ appropriately, we obtain the components \re{t14}. 
Two possibilities are\
a) $ H= H_6$ generated by $\s_{\ref{s68}}= (1\,2)(3\,6\,7\,4\,5\,8)$,
b)  $ H= H_8$ generated by the set
 $ \{ (1\,2)(3\,4)(5\,6)(7\,8) , (1\,3)(2\,4)(5\,7)(6\,8) , (1\,5)(2\,6)(3\,7)(4\,8)\}$.
\eex
The $\phi_{mnpq}$ obtained this way satisfy the well-known relations (e.g. \cite{cdfn}),
\begin{equation}
\phi_{ijk8} = \psi_{ijk}\ ,\quad
\phi_{ijkl} = \fr16 \e_{ijklmnp} \psi_{mnp}\ ,\quad  i,\dots,p=1,\dots,7\,.
\label{tspin7}
\end{equation}

\bex
Analogously to \re{D1}, we may directly obtain the Spin(7)-invariant set 
of $\phi$'s in \re{t14} from $\e_{ij}\ ,\ \e_{12}=1,$ in two dimensions:
\begin{equation}
\phi_{\s(i)\s(j)\s(7)\s(8)} = \e_{ij} \ ,\ 
                                    i,j=1,\dots,2\ ,\ 
                        \s\in  H_{21} \subset S_8\,,
\la{e2tot}
\end{equation}
where $ H= H_{21}$ is generated by $(3)(1264758)$ and $(1)(6)(234)(587)$.
\eex

\section{A  $\boldsymbol{D\!=\!10}$ structure  from a Spin(7) structure in
$\boldsymbol{d\!=\!8}$}

\subsection{Construction of a 6-form in $\boldsymbol{D\!=\!10}$ from a
Spin(7)-invariant 4-form in $\boldsymbol{d\!=\!8}$} 

Consider the Spin(7)-invariant self-dual 4-form
$\phi_{mnpq}$ in $d=8$  given in \re{t14}. Its discrete symmetry group
$G_{168}$ generated by the permutations \re{symm168} include the 
$\bZ_2\times \bZ_2$ transformations generated by the $2^4$ cycles
\begin{equation}
\rho_1 := (1 8)(2 7)(3 6)(4 5) \quad,\quad
\rho_2 := (1 4)(2 3)(5 8)(6 7)\ .
\la{reflections}
\end{equation}
Define the permutation $\s$ of $d$ indices, for $d$ even,
\begin{equation}
\s := (1 3 5 \dots d{-}1) (2 4 6 \dots d)\,,
\label{shift}
\end{equation}
which acts on the set of ordered pairs $\{1,2 \},\{3,4 \},\{5,6 \},\dots$ .
We see that for $d=8$, this mapping squared, $\s^2=\rho_1\cdot \rho_2$.

We want to embed the form $\phi$ \re{tspin7} into a form in $\bR^{10}$ 
with orthonormal basis  $(e_n)_{n=1,\dots,10}$. For the 
components, we shall denote the 10th index by a 0. Clearly, a 6-form $\Omega^0$
in ten dimensions which reduces to the above 4-form in  eight dimensions may
be defined in a trivial fashion by requiring the non-zero components
to be given by
\begin{equation}
\Omega^0_{mnpq\, 9\, 0} = \phi_{mnpq}\ ,\ m,n,p,q=1,\dots,8,
\label{Omega0}
\end{equation}
i.e.\ the 6-form $\Omega^0$  contracted with the volume form on the 9--10
plane yields the Spin(7)-invariant tensor \re{tspin7}.
However, there is a less trivial possibility.

Consider a 6-form $\Omega^1$ in $D=10$ with non-zero components
\begin{equation}
\Omega^1_{\s(m)\,\s(n)\,\s(p)\,\s(q)\, 1\, 2} = \phi_{mnpq}\ ,\ m,n,p,q=1,\dots,8.
\label{Omega1}
\end{equation}
We note that the components of $\Omega^1$ are compatible with the components
of $\Omega^0$, in that
\begin{equation}
\Omega^1_{mnpq\, 9\, 0} = \Omega^0_{mnpq\, 1\, 2}\ ,\ m,n,p,q=3,\dots,8\ .
\la{comp01}
\end{equation}
Similarly, the further 6-form having non-zero components
\begin{equation}
\Omega^2_{\s^2(m)\,\s^2(n)\,\s^2(p)\,\s^2(q)\, 3\, 4} = \phi_{mnpq}\ ,\
m,n,p,q\in \{1,\dots,8\}\backslash\{3,4\}
\label{Omega2}
\end{equation}
is consistent with both $\Omega^0$ and $\Omega^1$, i.e.
\bea
\Omega^0_{mnpq\, 3\, 4} &=& \Omega^2_{mnpq\, 1\, 2}\ ,\
m,n,p,q=5,\dots,10\nonumber\\
\Omega^1_{mnpq\, 3\, 4} &=& \Omega^2_{mnpq\, 9\,0}\ ,\
m,n,p,q=1,2,5,\dots,8.
\label{comp012}
\eea

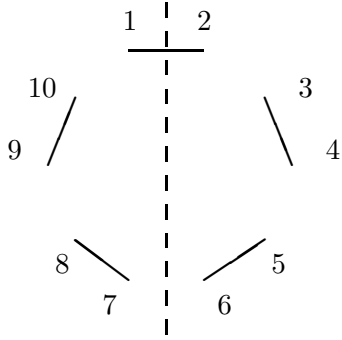
\begin{figure}
\setlength{\unitlength}{3729sp}%
\begingroup\makeatletter\ifx\SetFigFont\undefined%
\gdef\SetFigFont#1#2#3#4#5{%
  \reset@font\fontsize{#1}{#2pt}%
  \fontfamily{#3}\fontseries{#4}\fontshape{#5}%
  \selectfont}%
\fi\endgroup%
\begin{center}
\begin{picture}(2115,2249)(14626,-4178)
\thicklines
\put(15431,-3796){\line(-4, 3){356.800}}
\put(15931,-3796){\line( 3, 2){405}}
\put(16336,-2581){\line( 2,-5){180}}
\multiput(15681,-1951)(0.00000,-191.73913){12}{\line( 0,-1){ 95.870}}
\put(15431,-2266){\line( 1, 0){500}}
\put(14896,-3031){\line( 2, 5){180}}
\put(16741,-2986){\makebox(0,0)[lb]{\smash{\SetFigFont{11}{13.2}{\rmdefault}{\mddefault}{\updefault}4}}}
\put(16021,-4021){\makebox(0,0)[lb]{\smash{\SetFigFont{11}{13.2}{\rmdefault}{\mddefault}{\updefault}6}}}
\put(15391,-2131){\makebox(0,0)[lb]{\smash{\SetFigFont{11}{13.2}{\rmdefault}{\mddefault}{\updefault}1}}}
\put(15886,-2131){\makebox(0,0)[lb]{\smash{\SetFigFont{11}{13.2}{\rmdefault}{\mddefault}{\updefault}2}}}
\put(16561,-2581){\makebox(0,0)[lb]{\smash{\SetFigFont{11}{13.2}{\rmdefault}{\mddefault}{\updefault}3}}}
\put(16381,-3751){\makebox(0,0)[lb]{\smash{\SetFigFont{11}{13.2}{\rmdefault}{\mddefault}{\updefault}5}}}
\put(15256,-4021){\makebox(0,0)[lb]{\smash{\SetFigFont{11}{13.2}{\rmdefault}{\mddefault}{\updefault}7}}}
\put(14941,-3751){\makebox(0,0)[lb]{\smash{\SetFigFont{11}{13.2}{\rmdefault}{\mddefault}{\updefault}8}}}
\put(14761,-2581){\makebox(0,0)[lb]{\smash{\SetFigFont{11}{13.2}{\rmdefault}{\mddefault}{\updefault}10}}}
\put(14626,-2986){\makebox(0,0)[lb]{\smash{\SetFigFont{11}{13.2}{\rmdefault}{\mddefault}{\updefault}9}}}
\end{picture}\end{center}
%\vskip -1.5cm
\caption{The 6-form $\Omega$  is symmetric under $2\pi/5$ rotations generating
a $Z_5$ symmetry. It is antisymmetric under reflections in the dotted line.}
\label{fig}
\end{figure}

\noindent
In fact the five 6-forms
$\Omega^N_{\s^N(m)\,\s^N(n)\,\s^N(p)\,\s^N(q)\, \s^N(9)\,\s^N(0)}\,,\,
N=0,\dots,4$ are all compatible, allowing the definition of a 6-form
in ten dimensions manifestly invariant under the $\bZ_5$ transformations
between the five ordered pairs of indices in Figure \ref{fig} generated by
$\s= (13579)(24680)$, i.e.
\begin{equation}
\{1, 2\}\ra \{3 ,4 \}\ra \{ 5,6 \}\ra \{7 ,8 \}\ra \{9 ,10\}\ra\{1, 2\}\ .
\label{Z5}
\end{equation}
This $\bZ_5$-invariant 6-form has components given by
\bea
&& \Omega_{\,\s^N(m)\,\s^N(n)\,\s^N(p)\,\s^N(q)\,\s^N(9)\,\s^N(0)}
= \phi_{mnpq}\  ,\nonumber
\\  &&\hskip4cm  N=0,\dots,4\ ,\  m,n,p,q=1,\dots,8.
\label{spin7sixform}
\eea
Explicitly, for the choice \re{t14}, these are given by the 50 non-zero
elements
\begin{equation}
\hspace{-.05 cm}
\begin{array}{c c c c c c c c c c c c l}
\!\!\!\!&\ \Omega_{1 2 3 4 5 6} &\!\!=\!\!& \ \Omega_{1 2 3 4 7 8} 
&\!\!=\!\!& \ \Omega_{1 2 3 4 9 0}
&\!\!=\!\!
& \ \Omega_{1 2 3 5 7 9} &\!\!=\!\!&-\Omega_{1 2 3 5 8 0}
\cr 
\!\!=\!\!& -\Omega_{1 2 3 6 7 0} &\!\!=\!\!& -\Omega_{1 2 3 6 8 9}  &\!\!=\!\!& -\Omega_{1 2 4 5 7 0} 
&\!\!=\!\!&  -\Omega_{1 2 4 5 8 9}
 &\!\!=\!\!& -\Omega_{1 2 4 6 7 9} 
\cr 
\!\!=\!\!&\ \Omega_{1 2 4 6 8 0}
&\!\!=\!\!&   \ \Omega_{1 2 5 6 7 8} &\!\!=\!\!& \ \Omega_{1 2 5 6 9 0} &\!\!=\!\!& \ \Omega_{1 2 7 8 9 0}
&\!\!=\!\!& -\Omega_{1 3 4 5 7 9}
\cr
\!\!=\!\!&\Omega_{1 3 4 5 8 0}
&\!\!=\!\!&\ \Omega_{1 3 4 6 7 0} &\!\!=\!\!& \ \Omega_{1 3 4 6 8 9}
&\!\!=\!\!&\ \Omega_{1 3 5 6 7 9} &\!\!=\!\!& -\Omega_{1 3 5 6 8 0}
\cr 
\!\!=\!\!& -\Omega_{1 3 5 7 8 9} &\!\!=\!\!& \ \Omega_{1 3 5 7 9 0}
&\!\!=\!\!& \ \Omega_{1 3 6 7 8 0} &\!\!=\!\!&  -\Omega_{1 3 6 8 9 0}
&\!\!=\!\!& -\Omega_{1 4 5 6 7 0} 
\cr 
\!\!=\!\!& -\Omega_{1 4 5 6 8 9} &\!\!=\!\!& \ \Omega_{1 4 5 7 8 0}
&\!\!=\!\!& -\ \Omega_{1 4 5 8 9 0} &\!\!=\!\!&\ \Omega_{1 4 6 7 8 9} &\!\!=\!\!&  -\Omega_{1 4 6 7 9 0}
\cr
\!\!=\!\!&\ \Omega_{2 3 4 5 7 0}
&\!\!=\!\!& \ \Omega_{2 3 4 5 8 9}
&\!\!=\!\!&\ \Omega_{2 3 4 6 7 9} &\!\!=\!\!&  -\Omega_{2 3 4 6 8 0}
&\!\!=\!\!& -\Omega_{2 3 5 6 7 0} 
\cr 
\!\!=\!\!& -\Omega_{2 3 5 6 8 9}
&\!\!=\!\!&  \ \Omega_{2 3 5 7 8 0}
 &\!\!=\!\!& -\Omega_{2 3 5 8 9 0} &\!\!=\!\!& \ \Omega_{2 3 6 7 8 9} &\!\!=\!\!& -\Omega_{2 3 6 7 9 0}
\cr 
\!\!=\!\!& -\Omega_{2 4 5 6 7 9}  &\!\!=\!\!& \ \Omega_{2 4 5 6 8 0}
&\!\!=\!\!&\ \Omega_{2 4 5 7 8 9} &\!\!=\!\!&  -\Omega_{2 4 5 7 9 0}
&\!\!=\!\!& -\Omega_{2 4 6 7 8 0}
\cr 
\!\!=\!\!&\Omega_{2 4 6 8 9 0} &\!\!=\!\!&\ \Omega_{3 4 5 6 7 8}
&\!\!=\!\!& \ \Omega_{3 4 5 6 9 0} &\!\!=\!\!& \ \Omega_{3 4 7 8 9 0}
&\!\!=\!\!& \ \Omega_{5 6 7 8 9 0} &\!\!= 1\,.
\label{fifty}
\end{array}
\end{equation}
The vertex space $\cP^6(10)$ consists of $50$ points, corresponding to
a 49-valent graph having two types of vertices:  
Type A vertices connected to 30
vertices at distance 2,  16  vertices at distance 3 and 3 vertices at
distance 4 and  Type B vertices connected to 4 vertices at distance 1,
24 vertices at distance 2,  16  vertices at distance 3 and 5 vertices
at distance 4.  There are 10 vertices of Type A and 40 of Type B.

The symmetries of this democratic form are as follows. The permutation symmetry 
group is the order 60 alternating group  $A_5$ of five elements, the five ordered
pairs in \re{Z5}. 

The number of permutation antisymmetries is also 60, 
obtained from the elements of the permutation symmetry group by multiplication by,
for example, the reflection in the vertical axis of Figure \ref{fig}:
\begin{equation}
\tau:= (12)(03)(94)(85)(76)\ .
\label{antiperm6}
\end{equation}
There are 120 orthogonal symmetries and 120 orthogonal antiymmetries. 
The orthogonal symmetries which are not permutation symmetries have 
{\it all} their respective $\eta_i=-1$. 

In the eight dimensional subspaces orthogonal to nonexceptional planes $\{a,b\}$,
not in the set \re{Z5}, this 6-form reduces to an SU(2)-invariant 4-form, 
which we discuss further in section \ref{nonex}.

\subsection{Self-duality}

The six-form $\Omega$ given by \re{spin7sixform} defines skew-symmetric endomorphisms 
on the space of 3-forms, yielding generalised self-duality equations analogous 
to \re{tdual}
\begin{equation}
\fr16\ g^{m_4n_1}g^{m_5n_2}g^{m_6n_3} \Omega_{m_1m_2m_3m_4m_5m_6} G_{n_1n_2n_3} 
=  \lambda  G_{m_1m_2m_3}\ .
\la{gdual}
\end{equation}
Its  4-form dual $\F=\star \Omega$ defines a symmetric
endomorphism on the space of 2-forms, satisfying equations of the form
\re{tdual}. To find the eigenvalues of a $2k$-form $\Omega$ on $\LR{k}{D} $, we
identify the components of the $k$-forms $G$ in $D$ dimensions
\begin{equation}
\{G_{m_1m_2\dots m_k}\,, 1\le m_1 < m_2 < \dots < m_k \le D \}
\end{equation}
with the components of a vector in the $\tbinom{D}{k}$-dimensional space
$\LR{k}{D} $ thus:
\begin{equation}
G_A = G_{m_1m_2\dots m_k}\quad\mbox{where}\quad
  A= 1+ \sum_{i=1}^{k} \tbinom{m_i-1}{i} =1,\dots,\tbinom{D}{k}\ .
\end{equation}
On this vector the $2k$-form $\Omega$ may be represented as a
$\tbinom{D}{k}\times\tbinom{D}{k}$ matrix
\begin{equation}
\Omega_{AB}= \Omega_{m_1\dots m_k m_{k+1}\dots m_{2k}}\ ,\quad
A := 1+ \sum_{i=1}^{k} \tbinom{m_i-1}{i}\ ,\quad
B := 1+ \sum_{i=k+1}^{2k} \tbinom{m_i-1}{i}\ .
\end{equation}
In this notation, self-duality equations like \re{gdual} and \re{tdual}
take the form of matrix equations allowing direct evaluation of the
eigenvalues and eigenvectors using an algebraic computation programme like
Maple or Reduce.

We find the characteristic polynomials for the 6-form \re{spin7sixform} to be
\begin{equation}
\left(\lambda^6 + 51\lambda^4 + 699\lambda^2 + 1369\right)^4
\left( \lambda^4 + 42\lambda^2 + 361 \right)^6
\left( \lambda^2 + 1\right)^{35} \left( \lambda^2 + 9\right)
\label{charpolysix1}\end{equation}
and that of its dual 4-form to be
\begin{equation}
\left(\lambda+4\right) \left(\lambda+1\right)^8
\left(\lambda-1\right)^{24} \left( \lambda^2+2\lambda -19\right)^6 \ .
\end{equation}
We have checked that the stability group $H'\subset SO(10)$ of $\Omega$ 
(or equivalently $\F=\star \Omega$) has dimension 16 and is the direct product
\begin{equation}
H'=  SU(4)\ot  U(1)\ .
\label{stability}
\end{equation}
Under this stabilty group the $D=10$ dimensional vector module $V$
and the 45, 120 and 210 dimensional spaces of the two- three- and 
four-forms, respectively, decompose as
\bea
V \ =& {\bf 10}  &=\ \ 4_0 + {\ol 4}_0 + 1_{+1} + 1_{-1}
\\
\La^2 V \ =& {\bf 45} &=\  \ 15_0 + 6_0 + 6_0  + 4_{+1} + 4_{-1}
                                + {\ol 4}_{+1} + {\ol 4}_{-1} + 1_0 + 1_0
\nonumber\\
\La^3 V \ =& {\bf 120}  &=\ \
20_0 + \ol{20}_0 +  15_{+1} +  15_{-1} + (4_0)^3+ ({\ol 4}_0)^3 \nonumber\\
              &&\quad\ + (6_{+1})^2 + (6_{-1})^2 + 1_{+1}+ 1_{-1}
\nonumber\\
\La^4 V \ =& {\bf 210} &=\ \   20_0' + 20_{+1} + 20_{-1} + \ol{20}_{+1}
            + \ol{20}_{-1} + 15_0 + 15'_0 + 10_0 + \ol {10}_0
\nonumber\\  &&\quad\  + (6_0)^4 + (4_{+1})^2 + (4_{-1})^2
                 + ({\ol 4}_{+1})^2 + ({\ol 4}_{-1})^2 + (1_0)^4
\nonumber\eea
where the exponent denotes the multiplicity and the subscript the U(1)
charge. We identify the eigenspaces in $\La^2 V$ of the 4-form $\F$ as
follows:
\begin{equation}\arr
&\lambda =+1  &\Leftrightarrow  15  + 4 + {\ol 4} + 1\\
&\lambda =-1  &\Leftrightarrow   4 + {\ol 4} \\
&\lambda =-4  &\Leftrightarrow  1 \\
&\lambda =-1-2\sqrt{5} &\Leftrightarrow 6 \\
&\lambda =-1+2\sqrt{5} &\Leftrightarrow 6 \ .
\ea \end{equation}
The $\lambda =1$ eigenspace is the most interesting, satisfying a set of
21 equations amongst the 45 components of $\La^2 V$. The equations for the
other eigenspaces are rather overdetermined. For the six-form, the roots of
the 6th order polynomial in \re{charpolysix1} correspond to the  $4$'s and
${\ol 4}$'s, the roots of the quartic correspond to the four $6$'s,  the
eigenspaces with $\lambda =i$ and $-i$ transform as $(20 + 15)$ and
$(\ol{20} + 15)$, respectively, and the two singlets have eigenvalues $\pm
i\sqrt{3}$.

\subsection{Reduction to nonexceptional planes \la{nonex}}
As we have seen, in the five exceptional eight dimensional embeddings in
ten dimensions, the 6-form \re{fifty} reduces to the Spin(7)-invariant
4-form \re{t14}. Contracted with the bivectors  spanning 
all the other planes, i.e. for $\{a,b\}$ not in the set of planes in
\re{Z5}, we obtain a 4-form with  17 non zero components  $T_{pqrs}$. For
example in the space orthogonal to the $\{1,10\}$ plane we have
\begin{eqnarray}
&& T_{1 2 5 6} = - T_{1 6 7 8} = - T_{2 3 5 6} = -T_{1 3 5 7}
= - T_{3 4 6 7} = -T_{1 4 5 8} =  T_{2 4 5 7}
 \nonumber\\
&& = - T_{2 3 4 7} = T_{1 2 4 7} = - T_{2 5 6 7} =  T_{3 5 6 8}
= -T_{1 2 3 8} =  T_{2 5 7 8} =  T_{3 4 5 6}
 \nonumber\\
&&= -T_{2 4 6 8} =  T_{3 4 7 8} =  T_{1 3 4 6} = 1
\label{t17}
\end{eqnarray}
This 4-form in the eight dimensional space orthogonal to the $\{1,10\}$ plane
is invariant under an  SU(2) subgroup of SO(8). Under this SU(2) the
8-dimensional vector module of SO(8) decomposes as:
\begin{equation}
{\bf 8} =  4 [0] \op 2 [\fr12]\ ,
\label{eight}\end{equation}
i.e. four spin 0 modules and two spin $\fr12$ modules.
Here $[s]$ denotes the $2s+1$ dimensional module of SU(2).
The infinitesimal generators of $\su(2)$ are the following
$8\times 8$ matrices acting on the subspace $V^{2-9}$ spanned by
the basis vectors $e_2,\dots,e_9$:
\bea
T_1 &=& \frac{1}{2\sqrt 3}\begin{pmatrix}
                                   0 &  0 & 0 & 0 & 0 & 0 & 0 & 0 \cr
                                   0 &  0 & 0 & 1 & 0 &-1 & 0 & 0 \cr
                                   0 &  0 & 0 & 0 & 1 & 0 &-1 & 0 \cr
                                   0 & -1 & 0 & 0 & 0 & 1 & 0 & 0 \cr
                                   0 &  0 &-1 & 0 & 0 & 0 & 1 & 0 \cr
                                   0 &  1 & 0 &-1 & 0 & 0 & 0 & 0 \cr
                                   0 &  0 & 1 & 0 &-1 & 0 & 0 & 0 \cr
                                   0 &  0 & 0 & 0 & 0 & 0 & 0 & 0 
\end{pmatrix}
\nonumber\\[8pt]
T_2 &=& \frac{1}{2\sqrt 3} \begin{pmatrix}
                                   0 &  0 & 0 & 0 & 0 & 0 & 0 & 0 \cr
                                   0 &  0 & 0 & 0 & 1 & 0 &-1 & 0 \cr
                                   0 &  0 & 0 &-1 & 0 & 1 & 0 & 0 \cr
                                   0 &  0 & 1 & 0 &-1 & 0 & 0 & 0 \cr
                                   0 & -1 & 0 & 1 & 0 & 0 & 0 & 0 \cr
                                   0 &  0 &-1 & 0 & 0 & 0 & 1 & 0 \cr
                                   0 &  1 & 0 & 0 & 0 &-1 & 0 & 0 \cr
                                   0 &  0 & 0 & 0 & 0 & 0 & 0 & 0 
\end{pmatrix}
\nonumber\\[8pt]
T_3 &=&\ \frac{1}{6}\quad \begin{pmatrix}
                                   0 &  0 & 0 & 0 & 0 & 0 & 0 & 0 \cr
                                   0 &  0 & 2 & 0 &-1 & 0 &-1 & 0 \cr
                                   0 & -2 & 0 & 1 & 0 & 1 & 0 & 0 \cr
                                   0 &  0 &-1 & 0 &-1 & 0 & 2 & 0 \cr
                                   0 &  1 & 0 & 1 & 0 &-2 & 0 & 0 \cr
                                   0 &  0 &-1 & 0 & 2 & 0 &-1 & 0 \cr
                                   0 &  1 & 0 &-2 & 0 & 1 & 0 & 0 \cr
                                   0 &  0 & 0 & 0 & 0 & 0 & 0 & 0
\end{pmatrix}
\la{3matrices}
\eea
satisfying the standard commutation relations $[T_i,T_j]= \e_{ijk} T_k$.
These $T_i$'s clearly have nontrivial action on the subspace $V^{3-8}$ spanned
by the basis vectors $e_3,\dots,e_8$. In the eight dimensional space  $V^{2-9}$
the vectors\linebreak $(1,0,\dots,0), (0,\dots,0,1), (0,1,0,1,0,1,0,0)$ and
$(0,0,1,0,1,0,1,0)\ $ are the four invariant vectors under the $\su(2)$ action.
We find that the two spinor modules in \re{eight} are spanned by the basis
vectors
\begin{equation} \{ b_1{=} (0,1+\sqrt{3}i,0,1-\sqrt{3}i,0,-2,0,0) ,\
b_2{=} (0,0,\sqrt{3}+i,0,-\sqrt{3}+i,0,-2i,0)\}
\end{equation}
and
\begin{equation}
\{ c_1{=} (0,0,1+\sqrt{3}i,0,1-\sqrt{3}i,0,-2,0) ,\
c_2{=} (0,\sqrt{3}-i,0,2i,0,-\sqrt{3}-i,0,0)\}\ .
\end{equation}
$b_1$ and $c_1$ are the eigenvectors of $T_1$ with eigenvalue $-i/2$, whereas
$b_2$ and $c_2$ are its eigenvectors with eigenvalue $i/2$.

The above decomposition of the 8-dimensional vector module leads immediately
to the following decomposition of the 28-dimensional space of 2-forms
\begin{equation}
{\bf 28} = 9[0] \op  8 [\fr12] \op [1]\ .
\label{twoforms}
\end{equation}
With the 4-form $T$ defined in \re{t17}, the characteristic polynomial of the
self-duality equation \re{tdual} is
\begin{eqnarray}
\lambda^4 (\lambda^{2} -5)(\lambda^{4}  - 8\lambda^{2} + 3)^4 (\lambda^{6} -14\lambda^{4} + 33\lambda^{2} - 12)
\ . \label{char}
\end{eqnarray}

Let us now discuss the association of the roots of \re{char} with the
decomposition of the space of 2-forms \re{twoforms} under $\su(2)$:
\begin{itemize}
\item
We identify the eight roots of $(\lambda^{2} -5)(\lambda^{6} -14\lambda^{4} +33\lambda^{2} -12)$,
which are all distinct, with eight of the nine spin $[0]$ states in
\re{twoforms}.
\item
The four zero eigenvalues correspond to the ninth spin $[0]$
state together with the spin $[1]$ state.
\item
The remaining 16 eigenvalues,
the roots of $(\lambda^{4}  - 8\lambda^{2} + 3)^4$,
correspond to the spin [$\fr12$] states in the decomposition \re{twoforms}.
Since there are only four distinct eigenvalues $\pm\lambda_i\,,\, i=1,2$
the corresponding eigenspaces transform as the four dimensional
$[\fr12]\op[\fr12]$ representation.
\end{itemize}

\section{SU(4)$\ot$ U(1)-invariant 4-forms in ten dimensions}
As we see from the decomposition of the
${\bf 210}= \La^4(4_0 + \ol{4}_0 + 1_{+1} + 1_{-1})$, SU(4)$\ot$ U(1)
has four singlets.
One of them arises from the tensor product $4\ot\ol{4} \ot 1 \ot 1$
and the other three singlets have their origin in
$\La^4(4_0 + \ol{4}_0)$ and correspond to the three SU(4)-invariant 4-forms
discussed in the Appendix B of \cite{cdfn}. Choosing complex
coordinates $z_1 =x_1+ix_2\,,\,z_2 =x_3+ix_4\,,\,z_3 =x_5+ix_6\,,\,
z_4 =x_7+ix_8\,,\, z_5 =x_9+ix_{10}$, the four  SU(4)$\ot$ U(1)-invariant
forms may be expressed thus:
\bea
\F^{A}_{mnpq} &=& \sum_{1\le i< j < k\le 5}
\e_{mnpq z_i \bar{z}_i z_j \bar{z}_j z_k \bar{z}_k}
\label{inva}\\
\F^{B}_{mnpq}
&=& \sum_{\pi\in\bZ_5} \e_{mnpq z_1 \bar{z}_1 z_2 \bar{z}_2
            \left( z_3 \bar{z}_4 + z_4 \bar{z}_5 + z_5 \bar{z}_3 \right)}
    + \mbox{c.c.}
\label{invb}\\[8pt]
\F^{C}_{mnpq} + i\F^{D}_{mnpq} &=&
\e_{mnpq \bar{z}_1 \cdots \bar{z}_5 (z_1+\cdots +z_5)} \ ,
\label{invcd}
\eea
where the sum in $\F^B$ is over all cyclic permutations of $(1,\dots,5)$.
The corresponding vertex spaces $\cP^4(10)$ consist of $10$ points for $\F^{A}$,
$60$ points for $\F^{B}$ and $40$ points for $\F^{C}$ and $\F^{D}$.
The corresponding graphs have completely democratic vertices:  
For $\F^{A}$ every vertex is connected to 6 vertices at distance 2 and 
3  vertices at distance 4. For $\F^{B}$ every vertex is connected to 
6 vertices at distance 1, 27  vertices at distance 2, 30  vertices at distance 3
and 6  vertices at distance 4.  For $\F^{C}$ and $\F^{D}$  every vertex is 
connected to 4 vertices at distance 1, 18 vertices at distance 2,
12  vertices at distance 3 and 5  vertices at distance 4.

We find the characteristic polynomials of these 4-forms to be
\bea
\F^A &:& (\lambda - 1)^{20} (\lambda + 1 )^{24}(\lambda - 4)\\
\F^B &:&(\lambda+2)^{12} (\lambda-3)^{8} (\lambda-2)^{15} (\lambda+3)^{8}
         (\lambda^2+6\lambda-36) \\
\F^C,\F^D &:& \lambda^{33} (\lambda^2-20)^{6}\,.
\eea
The duals of these 4-forms yield invariant 6-forms $\Omega^A,\Omega^B,\Omega^C$ and
$\Omega^D$ having characteristic polynomials
\bea
\Omega^A &:& (\lambda^2+1)^{55} (\lambda^2+9)^{5} \\
\Omega^B &:& (\lambda^2+36)^{1} (\lambda^4+60\lambda^2+144)^{4}
          (\lambda^2+4)^{27} (\lambda^2+9)^{24} \\
\Omega^C,\Omega^D &:& \lambda^{80} (\lambda^2+20)^{20} \ .
\eea
The eigenvalues of the irreducible summands of these  4-forms and 6-forms
are tabulated in Tables \ref{eval4} and  \ref{eval6}.
The 4-forms $\F^C$ and $\F^D$ have identical eigenvalues. However, as
endomorphisms of 2-forms, these do not commute.

\begin{table}\begin{center}
\scriptsize{
\begin{tabular}{|c|c|c|c|c| c|c|c|}
\hline
$\La^2 V$={\bf 45} & $1_0$  & $1_0$ & $15_0$ &  $4_{+1}+\ol{4}_{+1}$
&$4_{-1}+\ol{4}_{-1}$ & $6_0$ & $6_0$\\ \hline
$\F^A$ &  $-1$ & 4  & $-1$ &   1 &  $-1$  &   1 &    1\\
$\F^B$ & $-3(1{+}\sqrt{5})$& $-3(1{-}\sqrt{5})$ & 2 & 3 & $-3$
& $-2$ &$-2$\\
$\F^C,\F^D$  & 0 & 0 & 0 & 0 & 0  & $2\sqrt{5}$ & $-2\sqrt{5}$ \\
\hline
\end{tabular}}
\caption{Eigenvalues of the invariant 4-forms on the irreducible summands of
$\La^2 V$}
\label{eval4}
\end{center}\end{table}

\begin{table}\begin{center}
\scriptsize{
\begin{tabular}{|c| c| c| c|  c| c| c| c| c|  c| c|}
\hline
$\La^3 V$={\bf 120}
&$20_0$& $\ol{20}_0$& $15_{\pm 1} $ & $4_0^2$ & $(\ol{4}_0)^2$ &$4_0$ &
$\ol{4}_0$  &$6_{+1}^2$ & $6_{-1}^2$ &$1_{\pm 1}$  \\   \hline
$\Omega^A$
& $i$ & $-i$ & $\pm i$&  $3i$ & $-3i$& $i$ & $-i$&  $i$&  $-i$&  $\pm3i$
\\
$\Omega^B$
& $3i$ &  $-3i$&  $\pm2i$& $i(3{\pm}\sqrt{21})$ & $-i(3{\pm}\sqrt{21})$ &
$3i$&  $-3i$&  $2i$&  $-2i$&  $\pm 6i$\\
$\Omega^C,\Omega^D$
& $i2 \sqrt{5}$ & $-i2 \sqrt{5}$ &  0&  0& 0& 0& 0& 0& 0& 0\\
\hline
\end{tabular}}
\caption{Eigenvalues of the invariant 6-forms on the irreducible summands of
$\La^3 V$}
\label{eval6}
\end{center}\end{table}

The linear combination $-\F^{A}-\F^{C}$ is the dual of the 6-form $\Omega$
constructed in \re{spin7sixform}.

The 4-form $\F^{A}$, is the dual of the 6-form constructed
in a similar fashion to  \re{spin7sixform} from the
$SU(4)\ot U(1)/\bZ_4$-invariant 4-form in eight dimensions discussed
in \cite{cdfn}:
\begin{equation}
T^{SU(4)\ot U(1)}_{mnpq}
= \sum_{1\le i< j \le 4} \e_{mnpq z_i \bar{z}_i z_j \bar{z}_j }\ ,\
m,n,p,q=1,\dots,8\ .
\label{d8su4u1}
\end{equation}
Using this to define a sixform as in \re{spin7sixform}, we obtain
$\Omega^A = \star \F^A$.
This clearly also has the special property that contracted with the volume
form on any of the five exceptional planes in Figure \ref{fig} yields the
eight dimensional 4-form \re{d8su4u1}.

$\F^{D}$ is not invariant under
the $\bZ_2$ transformation indicated on Figure \ref{fig}.
The four (SU(4)$\ot$ U(1))-invariants in ten dimensions are generated
by the four dimensional centraliser of SU(4)$\ot$ U(1) in GL(10,$\bR$).
In other words, acting on our invariant tensor  \re{spin7sixform}
with the four-parameter set of global GL(10,$\bR$) transformations which
commute with SU(4)$\ot$ U(1), yields the four-parameter set of
(SU(4)$\ot$ U(1))-invariants above.

The four-forms $\F^{A},\F^{B},\F^{C}$ and $\F^{D}$ have permutation symmetry groups of
order 240, 240, 120 and 120, respectively, all having $A_{60}$ as
commutator subgroups.  The number of permutation antisymmetries are
0,0,120,120, 
the number of orthogonal symmetries, 122880=120 $\cdot 2^{10}$, 960, 480, 240
and the number of orthogonal antisymmetries, 0, 0, 480 and 240, respectively.

\section{Comments}

Having seen that interesting 6-forms arise from the formula
\re{spin7sixform}, the question arises whether there is a similarly
constructed 8-form yielding interesting eigenvalue equations for 
a 4-form invariant under some subgroup $H''\subset$ SO(d+4).

The anwer is ``yes'' for $\Omega^A$ and $\Omega^B$: We may construct consistent 
$\bZ_6$-invariant  8-forms in 12 dimensions with components given by
\begin{equation}
\Psi_{\,\s^N(m)\,\s^N(n)\,\s^N(p)\,\s^N(q)\,\s^N(r)\,\s^N(s)\,
        \s^N(11)\,\s^N(12)}
= \Omega^I_{mnpqrs}\ ,\  N=0,\dots,5,
\label{8form}
\end{equation}
for  $I=A,B$.
The corresponding 4-form duals have characteristic polynomials:
\bea
\F^A &:&   (\lambda+1)^{35} (\lambda-5) (\lambda-1)^{30}
\nonumber\\
\F^B &:&  (\lambda-2)^{24} (\lambda+2)^{20} (\lambda-4)^{10}
             (\lambda+4)^{10} (\lambda^2 + 8\lambda-80)\ .
\eea

\section*{Acknowledgements}
One of us (J.N.) thanks the Belgian Fonds National de la Recherche Scientifique
for travel support, the Max-Planck-Institut f\"ur Mathematik in Bonn as well as 
Prof.  Hermann Nicolai and the 
Max-Planck-Institut f\"ur Gravitationsphysik in Potsdam for hospitality.

%%%%%%%%%%%%%%%%%%%%%%%%%

\appendix

\section{Further results: SO($d$)-invariants}

We denote by $I_n\ ,\ n=1,\dots,N(p,d)$, the independent 
SO($d,\bR$)-invariants constructed from a $p$-form $\f$, 
where $N(p,d)$ is the number of invariants.

\bconj
The SO($d,\bR$) orbit of a special $p$-form is characterised
by the values of the invariants. 
\econj
This means that
a general $p$-form is special if and only if the values of the invariants 
are equal to the values of the invariants
of a representative special $p$-form, in whose orbit
it then lies. If two $p$-forms have identical invariants, they belong to the same 
SO($d,\bR$)-orbit.

As an example we list in Table \ref{p2d4canon} the complete set of 
representative special $2$-forms in 4-dimensions. The values of the two independent 
invariants 
\bea
I_1(\f)&=&\sum_{a,b} \f_{ab} \f_{ba}
     \nonumber\\
I_2(\f)&=&\sum_{a,b,c,d} \epsilon_{abcd}\f_{ab} \f_{cd}
\label{invp2d4}
\eea
are given.
If an arbitrary $2$-form in 4-dimensions
has values of $I_1,I_2$ appearing in the table, then 
it is in the orbit of the corresponding representative.

\begin{table}[ht]
\caption{Representative special 2-forms in 4 dimensions. The forms marked with D are
democratic. 
\la{p2d4canon}}
\vspace{1cm}
\hspace{2cm}
{\begin{tabular}
{|c  ||c   |c   |c   |   c   |c   |c   ||c   |c   |c   |c   |c   | }
\hline
\multicolumn{10}{|c|}{Representative forms for $p$=2, $d$=4}\\
\hline
&$\e_{12}$&$\e_{13}$&$\e_{14}$&$\e_{23}$&$\e_{24}$&$\e_{34}$
&$I_1$&$I_2$&   \\
\hline\hline
$A$ & 1 & 0 &0&0&0&0
&-2&0     &\\
\hline 
\hline 
$B_1$& 1 & 1  &0&0&0&0
&-4&0     &\\
\hline 
$B_2$& 1 & 0 &0&0&0&1
&-4&8    &D\\
\hline 
$B_3$& 1 & 0 &0&0&0&-1
&-4&-8    &D\\
\hline 
\hline 
$C_1$& 1 & 1 &1&0&0&0
&-6&0    &\\
\hline 
$C_2$& 1 & 1 &0&0&-1 &0
&-6& 8     &\\
\hline 
$C_3$& 1 & 1 &0&0&1 &0
&-6& -8    &\\
\hline 
\hline 
$D_1$& 1 & 1 & 1&1 &0&0
&-8& 8   &\\
\hline 
$D_2$& 1 & 1 & 1&-1 &0&0
&-8& -8  &\\
\hline 
$D_3$&1&1&0&0&-1&1
&-8& 16    &D\\ 
\hline 
$D_4$&1&1&0&0&1&1
&-8& 0    &D\\ 
\hline
$D_5$&1&1&0&0&1&-1
&-8& -16    &D\\ 
\hline 
\hline
$E_1$&1&1&1&1&-1&0
&-10& 16     &\\
\hline
$E_2$&1&1&1&1&1&0
&-10& 0     &\\
\hline
$E_3$&1&1&1&-1&1&0
&-10& -16     &\\
\hline 
\hline
$F_1$&1&1&1&1&-1&1
&-12& 24   &D\\
\hline
$F_2$&1&1&1&1&1&1
&-12& 8   &D\\
\hline
$F_3$&1&1&1&-1&-1&-1
&-12& -8   &D\\
\hline
$F_4$&1&1&1&-1&1&-1
&-12& -24   &D\\
\hline\hline
\end{tabular}}
\end{table}

%%%%%%%%%%%%%%%%%%%%%%%%%%%%%%%

\end{document}